\documentclass[12pt]{iopart}

\usepackage{iopams} 
\usepackage{citesort}

\usepackage{graphicx}

\begin{document}

\title{Can thermal emission from time-varying media be described semiclassically?}

\author{Iñigo Liberal, J.E. Vázquez-Lozano and \\ Antonio Ganfornina-Andrades}

\address{Department of Electrical, Electronic and Communications Engineering, \\ Institute of Smart Cities (ISC), Public University of Navarre (UPNA), \\ 31006 Pamplona, Spain}
\ead{inigo.liberal@unavarra.es}
\vspace{10pt}

\begin{abstract}
Time-varying media, i.e., materials whose properties dynamically change in time, have opened new possibilities for thermal emission engineering by lifting the limitations imposed by energy conservation and reciprocity, and providing access to nonequilibrium dynamics. In addition, quantum effects, such as vacuum amplification and emission at zero temperature, have been predicted for time-varying media, reopening the debate on the quantum nature of thermal emission. Here, we derive a semiclassical theory to thermal emission from time-varying media based on fluctuational electrodynamics, and compare it to the quantum theory. Our results show that a quantum theory is needed to correctly capture the contribution from quantum vacuum amplifications effects, which can be relevant even at room temperature and mid-infrared frequencies. Finally, we propose corrections to the standard semiclassical theory that enable the prediction of thermal emission from time-varying media with classical tools. 
\end{abstract}

\vspace{2pc}

\section{Introduction}

Thermal emission is one of the most fundamental and ubiquitous radiative processes, and it underpins photonic technologies such as radiative cooling \cite{Raman2014passive}, thermophotovoltaics \cite{Burger2020present}, thermal imaging \cite{Bao2023heat} and thermal camouflage \cite{Qu2018thermal}, novel light sources \cite{Ilic2016tailoring} and sensing \cite{Lochbaum2017chip}. For this reason, recent decades have witnessed very intensive research on the design of photonic nanostructures to enhance the spatial and temporal coherence of thermal fields, thus facilitating their control and manipulation, and the subsequent design of more efficient thermal emitters \cite{Baranov2019nanophotonic,Boriskina2017heat}. 

Recently, a new approach to engineer thermal emission has emerged based on the use of temporal metamaterials  and/or time-varying media \cite{Vazquez2023incandescent,Yu2023manipulating,Yu2023time,Buddhiraju2020photonic}, i.e., materials whose properties are dynamically modulated in time. Using time as an additional degree of freedom \cite{Engheta2023four,Galiffi2022photonics,Caloz2022generalized} fundamentally changes the physics of thermal emission, since breaking temporal symmetries lifts the restrictions imposed by energy conservation and reciprocity \cite{Ortega2023tutorial}. Moreover, temporal modulation moves the system away from thermal equilibrium, opening up new non-equilibrium regimes for thermal nanophotonic engineering. As a consequence, fluctuating currents in a time-modulated body exhibit nontrivial correlations in space, frequency and polarization \cite{Vazquez2023incandescent}. 

Within this context, epsilon-near-zero (ENZ) media \cite{Liberal2017near,Lobet2023new}, i.e., materials with a near-zero permittivity, are increasingly becoming a driving force in the field. First, the strong and ultrafast nonlinearities of ENZ media \cite{Alam2016large,Reshef2019nonlinear} provide a mechanism in which to implement time-varying media concepts. In fact, ENZ media based on doped semiconductors have facilitated the demonstration of time-refraction \cite{Zhou2020broadband}, spatiotemporal refraction \cite{Bohn2021spatiotemporal}, saturable mirrors \cite{Tirole2022saturable}, interference in temporal slits \cite{Tirole2023double} and single-cycle modulations \cite{Lustig2023time}. Second, bulk ENZ samples are characterized by intrinsically strong thermal fluctuating fields \cite{Liberal2018manipulating,Liberal2017zero}, which can be freed by means of time modulation \cite{Vazquez2023incandescent}. The resulting thermal emitter is the dual frequency momentum of conventional gratings, with near-to-far field partially coherent emission \cite{Vazquez2023incandescent}. 

We note that thermal emission from time-varying media has been derived from first principles based on a quantum electrodynamics approach \cite{Vazquez2023incandescent}, as well as with semiclassical theories based on fluctuational electrodynamics \cite{Yu2023manipulating,Yu2023time,Buddhiraju2020photonic}. Interestingly, the quantum character of thermal emission has been the subject of debate and interest for over a century. In fact, Planck’s thermal radiation law for correcting the ultraviolet catastrophe was the first motivation for quantizing the electromagnetic field \cite{Loudon2000quantum}. Similarly, photon bunching at the Hansbury Brown and Twiss (HBT) effect \cite{Brown1956correlation} opened the field of measuring photon statistics, and marked the onset of photon counting experiments and quantum optics as it is known today \cite{Glauber1963quantum}. However, nowadays both effects can be reproduced with classical theories \cite{Loudon2000quantum,Boyer1969derivation,Mandel1995optical}. Thus, while quantum optics has become a thriving field \cite{o2009photonic,Wang2020integrated}, semiclassical theories are most commonly employed to model thermal emission. Indeed, most of the contemporary thermal emitters are routinely modeled with the very successful theory of fluctuational electrodynamics \cite{Joulain2005surface,Rytov1989principles}. Despite this fact, the strong quantum effects (including quantum vacuum amplification and emission at zero temperature) predicted for time-modulated bodies pose questions on the need to correctly capture the quantum character of thermal emission in the case of time-varying media.
 
Here, we provide a discussion on the quantum character of thermal emission from time-varying media. To this end, we develop a semiclassical theory of thermal emission from time-varying media based on the usual framework of fluctuational electrodynamics. Moreover, the semiclassical theory is formulated in a way that enables a direct comparison with the quantum theory. In this manner, we are able to clarify when and why a quantum theory of thermal emission is needed for time-varying media.

\section{Theoretical framework}

\begin{figure}[t!]
\centering
\includegraphics[width=0.5\textwidth]{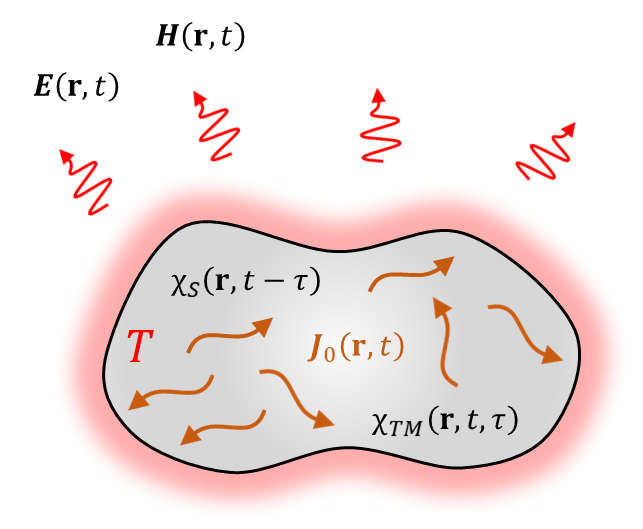}
\caption{ { \bf Semiclassical description of thermal emission from time-varying media.} A macroscopic body at temperature $T$, whose electromagnetic response is characterized by time-invariant 
$\boldsymbol{\chi}_{S}\left(\mathbf{r},t-\tau\right)$ and time-modulated
$\boldsymbol{\chi}_{TM}\left(\mathbf{r},t,\tau\right)$
dielectric susceptibility tensors, supports fluctuating currents
$\mathcal{\boldsymbol{J}}_{0}\left(\mathbf{r},t\right)$, resulting in the radiation of a classical (thermal) electromagnetic field $\left(\mathcal{\boldsymbol{E}}\left(\mathbf{r},t\right),\mathcal{\boldsymbol{H}}\left(\mathbf{r},t\right)\right)$.}
\label{fig:Sketch}
\end{figure}

As schematically depicted in Fig.\,\ref{fig:Sketch}, we consider thermal emission from a macroscopic body initially at thermal equilibrium at temperature $T$, whose material properties are modulated in time. Within a macroscopic theory, the material response of the body can be described via a time-invariant dielectric susceptibility tensor, 
$\boldsymbol{\chi}_{S}\left(\mathbf{r},t-\tau\right)$, and time-modulated dielectric susceptibility tensor 
$\boldsymbol{\chi}_{TM}\left(\mathbf{r},t,\tau\right)$. Being a semiclassical theory, electric 
$\mathcal{\boldsymbol{E}}\left(\mathbf{r},t\right)$ and magnetic $\mathcal{\boldsymbol{H}}\left(\mathbf{r},t\right)$ fields are considered as classical fields excited by current densities 
$\mathcal{\boldsymbol{J}}_{0}\left(\mathbf{r},t\right)$, whose statistical properties are justified within the framework of fluctuational electrodynamics \cite{Joulain2005surface,Rytov1989principles}. The total time-domain electric $\mathcal{\boldsymbol{E}}\left(\mathbf{r},t\right)$ and magnetic $\mathcal{\boldsymbol{H}}\left(\mathbf{r},t\right)$ fields in the system are found as the solution to Maxwell equations (MEs):

\begin{equation}
\nabla\times\mathcal{\boldsymbol{E}}\left(\mathbf{r},t\right)=
-\mu_{0}\,\,\partial_{t}\mathcal{\boldsymbol{H}}\left(\mathbf{r},t\right)
\label{eq:ME_curlE_time_domain}
\end{equation}

\[
\nabla\times\mathcal{\boldsymbol{H}}\left(\mathbf{r},t\right)
=\varepsilon_{0}\,\,\partial_{t}\mathcal{\boldsymbol{E}}\left(\mathbf{r},t\right)+\varepsilon_{0}\,\,\partial_{t}\int_{-\infty}^{\infty}d\tau\,\boldsymbol{\chi}_{S}\left(\mathbf{r},t-\tau\right)\mathcal{\boldsymbol{E}}\left(\mathbf{r},\tau\right)
\]

\begin{equation}
+\varepsilon_{0}\,\,\partial_{t}\int_{-\infty}^{\infty}d\tau\,\boldsymbol{\chi}_{TM}\left(\mathbf{r},t,\tau\right)\mathcal{\boldsymbol{E}}\left(\mathbf{r},\tau\right)
+\mathcal{\boldsymbol{J}}_{0}\left(\mathbf{r},t\right)
\label{eq:eq:ME_curlH_time_domain}
\end{equation}

Next, we transform the fields and currents to the frequency domain with Fourier transform pairs defined following the $e^{-i\omega t}$ time-harmonic convention, leading to MEs in the frequency domain

\begin{equation}
\nabla\times\mathbf{E}\left(\mathbf{r},\omega\right)=
i\omega\mu_{0}\,\,\mathbf{H}\left(\mathbf{r},\omega\right)
\label{eq:ME_curlE_freq}
\end{equation}

\[
\nabla\times\mathbf{H}\left(\mathbf{r},\omega\right)=
-i\omega\varepsilon_{0}\,\,\mathbf{E}\left(\mathbf{r},\omega\right)
-i\omega\varepsilon_{0}\boldsymbol{\chi}_{S}\left(\mathbf{r},\omega\right)\cdot\mathbf{E}\left(\mathbf{r},\omega\right)
\]

\begin{equation}
-i\omega\varepsilon_{0}\,\,\int_{-\infty}^{\infty}\frac{d\omega'}{2\pi}\,\,\boldsymbol{\chi}_{TM}\left(\mathbf{r},\omega,-\omega'\right)\cdot
\mathbf{E}\left(\mathbf{r},\omega'\right)+\mathbf{J}_{0}\left(\mathbf{r},\omega\right)
\label{eq:ME_curlH_freq}
\end{equation}

\noindent where we highlight that the time-dependent response of the system is characterized by a two-frequency response $\boldsymbol{\chi}_{TM}\left(\mathbf{r},\omega,-\omega'\right)$, while the non-time-modulated susceptibility is characterized by a single-frequency response, usually associated with the permittivity: 
$\boldsymbol{\varepsilon}_S(\mathbf{r},\omega)=
\varepsilon_0\,\,(\mathbf{I}+\boldsymbol{\chi}_{S}(\mathbf{r},\omega))$.

\begin{figure}[t!]
\centering
\includegraphics[width=0.7\textwidth]{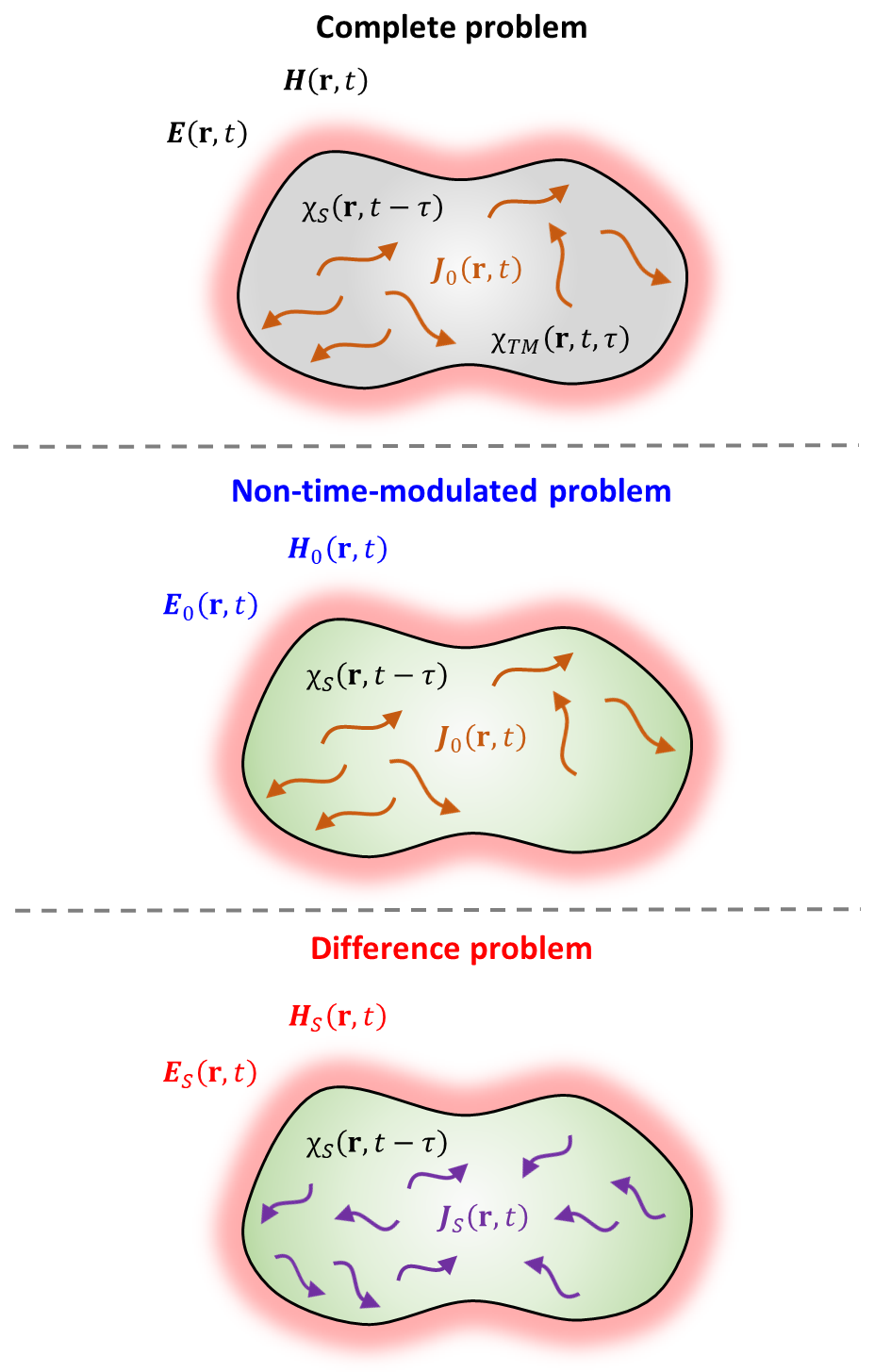}
\caption{ { \bf Volume equivalence theorem in time-varying media.} (Top) Original problem of emission from time-varying media. (Center) Auxiliary problem consisting of the same configuration, but in the absence of time-modulation. (Bottom) Difference problem recasted as a non-time-modulated problem, but in the presence of volumetric currents.}
\label{fig:Volume_equivalence}
\end{figure}

\section{Volume equivalence theorem}

Similar to electromagnetic scattering problems \cite{Balanis2012advanced}, we conveniently rewrite the problem by using an extension of the volume equivalence theorem to time-modulated media. To this end, we start by defining an auxiliary problem in the absence of time-modulation, whose electric $\mathbf{E}_{0}\left(\mathbf{r},\omega\right)$ and magnetic $\mathbf{H}_{0}\left(\mathbf{r},\omega\right)$ fields are found as the solutions to time-harmonic MEs in the absence of time-modulation (see Fig.\,\ref{fig:Volume_equivalence}):

\begin{equation}
\nabla\times\mathbf{E}_{0}\left(\mathbf{r},\omega\right)=i\omega\mu_{0}\mathbf{H}_{0}\left(\mathbf{r},\omega\right)
\label{eq:curl_E0}
\end{equation}

\begin{equation}
\nabla\times\mathbf{H}_{0}\left(\mathbf{r},\omega\right)=-i\omega\varepsilon_{0}\mathbf{E}_{0}\left(\mathbf{r},\omega\right)-i\omega\varepsilon_{0}\boldsymbol{\chi}_{S}\left(\mathbf{r},\omega\right)\cdot\mathbf{E}_{0}\left(\mathbf{r},\omega\right)+\mathbf{J}_{0}\left(\mathbf{r},\omega\right)
\label{eq:curl_H0}
\end{equation}

We also define the ``difference'' fields, which represent the fields radiated by the time-varying polarization currents, corresponding to how the field distributions change between both time-modulated and non-time-modulated problems

\begin{equation}
\mathbf{E}_{S}\left(\mathbf{r},\omega\right)
=\mathbf{E}\left(\mathbf{r},\omega\right)-\mathbf{E}_{0}\left(\mathbf{r},\omega\right)
\label{eq:ES}
\end{equation}

\begin{equation}
\mathbf{H}_{S}\left(\mathbf{r},\omega\right)
=\mathbf{H}\left(\mathbf{r},\omega\right)-\mathbf{H}_{0}\left(\mathbf{r},\omega\right)
\label{eq:HS}
\end{equation}

Importantly, by combining (\ref{eq:ME_curlE_freq})-(\ref{eq:ME_curlH_freq}) and (\ref{eq:curl_E0})-(\ref{eq:curl_H0}), it can be demonstrated that the difference fields obey
the following equations

\begin{equation}
\nabla\times\mathbf{E}_{S}\left(\mathbf{r},\omega\right)=i\omega\mu_{S}\mathbf{H}_{S}\left(\mathbf{r},\omega\right)
\label{eq:curlES}
\end{equation}

\begin{equation}
\nabla\times\mathbf{H}_{S}\left(\mathbf{r},\omega\right)=
-i\omega\varepsilon_{S}\,\,\mathbf{E}_{S}\left(\mathbf{r},\omega\right)
-i\omega\varepsilon_{S}\,\,\boldsymbol{\chi}_{S}\left(\mathbf{r},\omega\right)\cdot\mathbf{E}_{S}\left(\mathbf{r},\omega\right)
+\mathbf{J}_{S}\left(\mathbf{r},\omega\right)
\label{eq:curlHS}
\end{equation}

\noindent where we have defined

\begin{equation}
\mathbf{J}_{S}\left(\mathbf{r},\omega\right)=
-i\omega\varepsilon_{0}\,\,\int \frac{d\omega'}{2\pi}\,\,\boldsymbol{\chi}_{TM}\left(\mathbf{r},\omega,-\omega'\right)\cdot\mathbf{E}\left(\mathbf{r},\omega'\right)
\label{eq:J_S}
\end{equation}

In other words, the difference fields are the solutions to MEs in the non-time-modulated problem, but excited with a volumetric current distribution $\mathbf{J}_{S}\left(\mathbf{r},\omega\right)$, which represents the fields radiated by the polarization induced by the time-dependent polarizability (hence the volume equivalence theorem). 

Recasting this problem in such a way is convenient, since the solution to the non-time-modulated problem can be written as a function of a standard dyadic Green's function:

\begin{equation}
\mathbf{E}_{0}\left(\mathbf{r},\omega\right)=
i\omega\mu_{0}\,\int dV'\,\,\,
\mathbf{G}\left(\mathbf{r},\mathbf{r}',\omega\right)\cdot\mathbf{J}_{0}\left(\mathbf{r}',\omega\right)
\label{eq:E0_Green}
\end{equation}

\begin{equation}
\mathbf{E}_{S}\left(\mathbf{r},\omega\right)=
i\omega\mu_{0}\,\int dV'\,\,\,
\mathbf{G}\left(\mathbf{r},\mathbf{r}',\omega\right)\cdot\mathbf{J}_{S}\left(\mathbf{r}',\omega\right)
\label{eq:ES_Green}
\end{equation}

\noindent where $\mathbf{G}\left(\mathbf{r},\mathbf{r}',\omega\right)$ is the dyadic Green's function of the non-time-modulated problem, defined as the solution to
\begin{equation}
\nabla\times\nabla\times\mathbf{G}\left(\mathbf{r},\mathbf{r}',\omega\right)
-\frac{\omega^{2}}{c^{2}}\left(\mathbf{I}+\boldsymbol{\chi}_{S}\left(\mathbf{r},\omega\right)\right)\cdot\mathbf{G}\left(\mathbf{r},\mathbf{r}',\omega\right)
=\mathbf{I}\,\delta\left(\mathbf{r}-\mathbf{r}'\right)
\label{eq:Green_wave_equation}
\end{equation}

\noindent where $c=1/\sqrt{\varepsilon_0\mu_0}$ stands for the speed of light in vacuum.

An additional consequence of Eqs.~(\ref{eq:E0_Green})-(\ref{eq:ES_Green}) is that the total field emitted from the time-varying body can also be written as a function of the dyadic
Green's function acting on the total current: 
$\mathbf{J}\left(\mathbf{r},\omega\right)=\mathbf{J}_{0}\left(\mathbf{r},\omega\right)+\mathbf{J}_{S}\left(\mathbf{r},\omega\right)$, i.e., 

\begin{equation}
\mathbf{E}\left(\mathbf{r},\omega\right)
=i\omega\mu_{0}\,\int dV'\,\,\,\mathbf{G}\left(\mathbf{r},\mathbf{r}',\omega\right)\cdot\mathbf{J}\left(\mathbf{r}',\omega\right)
\label{eq:E_Green}
\end{equation}

The advantage of writing all field quantities in terms of the dyadic Green's function for a non-time-modulated configuration is that $\mathbf{G}\left(\mathbf{r},\mathbf{r}',\omega\right)$ is linear in frequency and significantly simpler than the dyadic Green's function for the time-modulated problem. Indeed, there is a large body of literature devoted to writing dyadic Green's functions in closed form, as well as to their numerical evaluation \cite{tai1971dyadic,Felsen1994radiation,Sipe1987new,Francoeur2009solution,Narayanaswamy2014green}. Therefore, a theoretical framework based on non-time-modulated dyadic Green's functions can be implemented in practice for a large number of configurations. In addition, we will show that this theoretical framework allows for a direct comparison with the quantum theory \cite{Vazquez2023incandescent}.

Finally, by introducing (\ref{eq:J_S}) into (\ref{eq:E_Green}) we can write the following implicit equation for the total field
\begin{equation}
\mathbf{E}\left(\mathbf{r},\omega\right)=\mathbf{E}_{0}\left(\mathbf{r},\omega\right)
+\frac{\omega^{2}}{c^2}\int dV'\,\int \frac{d\omega'}{2\pi}\,
\mathbf{G}\left(\mathbf{r},\mathbf{r}',\omega\right)\cdot\boldsymbol{\chi}_{TM}\left(\mathbf{r}',\omega,-\omega'\right)\cdot\mathbf{E}\left(\mathbf{r}',\omega'\right)
\label{eq:Ew_implicit}
\end{equation}

Inversely, by introducing (\ref{eq:E_Green}) into (\ref{eq:J_S}) we can write an implicit equation for the current density distribution
\begin{equation}
\mathbf{J}\left(\mathbf{r},\omega\right)=\mathbf{J}_{0}\left(\mathbf{r},\omega\right)+\int dV'\,\int \frac{d\omega'}{2\pi}\,\frac{\omega\omega'}{c^2}\,\boldsymbol{\chi}_{TM}\left(\mathbf{r},\omega,-\omega'\right)\cdot\mathbf{G}\left(\mathbf{r},\mathbf{r}',\omega'\right)\cdot\mathbf{J}\left(\mathbf{r}',\omega'\right)
\label{eq:Jw_Implicit}
\end{equation}

\section{Iterative solutions}

In the previous section, the solution for the radiation from a time-modulated system has been recasted in the form of integral equations (\ref{eq:Ew_implicit})-(\ref{eq:Jw_Implicit}) very similar to those use in classical computational electromagnetics \cite{Balanis2016antenna}, which might help in their numerical evaluation in a number of general cases. At the same time, it is expected that the time modulated susceptibility will be small at optical frequencies \cite{Saha2023photonic,Boyd_book}, which opens up a number of theoretical techniques for the analysis of thermal emission from time-varying media. For example, an iterative solution to the total field $\mathbf{E}\left(\mathbf{r},\omega\right)$ can be written in the following manner

\begin{equation}
\mathbf{E}\left(\mathbf{r},\omega\right)=\sum_{n=0}^{\infty}\,\,\,\mathbf{E}_{n}\left(\mathbf{r},\omega\right)
\label{eq:Ew_series}
\end{equation}

\[
=\mathbf{E}_{0}\left(\mathbf{r},\omega\right)
\]

\[
+\frac{\omega^{2}}{c^2}\int dV'\,\int \frac{d\omega'}{2\pi}\,\mathbf{G}\left(\mathbf{r},\mathbf{r}',\omega\right)\cdot\boldsymbol{\chi}_{TM}\left(\mathbf{r}',\omega,-\omega'\right)\cdot\mathbf{E}_{0}\left(\mathbf{r}',\omega'\right)
\]

\[
+\int dV'\,\int dV''\,\int \frac{d\omega'}{2\pi}\,\int \frac{d\omega''}{2\pi}\,
\frac{\omega^{2}}{c^2}\frac{\omega'^{2}}{c^2}
\mathbf{G}\left(\mathbf{r},\mathbf{r}',\omega\right)\cdot
\]

\[
\boldsymbol{\chi}_{TM}\left(\mathbf{r}',\omega,-\omega'\right)\cdot\mathbf{G}\left(\mathbf{r}',\mathbf{r}'',\omega'\right)\cdot\boldsymbol{\chi}_{TM}\left(\mathbf{r}'',\omega',-\omega''\right)\cdot\mathbf{E}_{0}\left(\mathbf{r}'',\omega''\right)
\]

\begin{equation}
+...
\end{equation}

In the end, one obtains a series solution that, if convergent for a sufficiently small time-dependent susceptibility, provides a solution to the problem. In practice, retaining a few terms of the series can be sufficient for a range of experimental configurations. The same process can be carried out for the current density distribution

\begin{equation}
\mathbf{J}\left(\mathbf{r},\omega\right)=\sum_{n=0}^{\infty}\mathbf{J}_{n}\left(\mathbf{r},\omega\right)
\end{equation}

\[
=\mathbf{J}_{0}\left(\mathbf{r},\omega\right)
\]

\[
+\int dV'\,\int_{-\infty}^{\infty}\frac{d\omega'}{2\pi}\,\frac{\omega}{c}\frac{\omega'}{c}\boldsymbol{\chi}_{TM}\left(\mathbf{r},\omega,-\omega'\right)\cdot\mathbf{G}\left(\mathbf{r},\mathbf{r}',\omega'\right)\cdot\mathbf{J}_{0}\left(\mathbf{r}',\omega'\right)
\]

\[
+\int dV'\,\int dV''\,\int_{-\infty}^{\infty}\frac{d\omega'}{2\pi}\,
\int_{-\infty}^{\infty}\frac{d\omega''}{2\pi}\,
\frac{\omega}{c}\left(\frac{\omega'}{c}\right)^{2}\frac{\omega''}{c}\,\,
\boldsymbol{\chi}_{TM}\left(\mathbf{r},\omega,-\omega'\right)\cdot
\]

\[
\mathbf{G}\left(\mathbf{r},\mathbf{r}',\omega'\right)\cdot\boldsymbol{\chi}_{TM}\left(\mathbf{r}',\omega',-\omega''\right)\cdot\mathbf{G}\left(\mathbf{r}',\mathbf{r}'',\omega''\right)\cdot\mathbf{J}_{0}\left(\mathbf{r}'',\omega''\right)
\]

\begin{equation}
+...
\end{equation}

As it will be useful for the derivation of the fluctuation-dissipation theorem, we next assume that the time-varying susceptibility is isotropic, 
$\boldsymbol{\chi}_{TM}\left(\mathbf{r}',\omega',-\omega''\right)=\mathbf{I}\,\,\chi_{TM}\left(\mathbf{r}',\omega',-\omega''\right)$, and write the iterative solution for individual current components

\begin{equation}
J_{p}\left(\mathbf{r},\omega\right)=\sum_{n=0}^{\infty}J_{p,n}\left(\mathbf{r},\omega\right)
\label{eq:Jp_series}
\end{equation}

\begin{equation}
J_{p,1}\left(\mathbf{r},\omega\right)=
\sum_{p'}\int dV'\,\int \frac{d\omega'}{2\pi}\,
\frac{\omega}{c}\frac{\omega'}{c}\,\chi_{TM}\left(\mathbf{r},\omega,-\omega'\right)G_{pp'}\left(\mathbf{r},\mathbf{r}',\omega'\right)J_{p',0}\left(\mathbf{r}',\omega'\right)
\label{eq:Jp_1}
\end{equation}

\[
J_{p,2}\left(\mathbf{r},\omega\right)=
\sum_{p',p''}\int dV'\,\int dV''\,\int \frac{d\omega'}{2\pi}\,\int \frac{d\omega''}{2\pi}\,
\frac{\omega}{c}\left(\frac{\omega'}{c}\right)^{2}\frac{\omega''}{c}
\]

\begin{equation}
\chi_{TM}\left(\mathbf{r},\omega,-\omega'\right)\chi_{TM}\left(\mathbf{r}',\omega',-\omega''\right)G_{pp'}\left(\mathbf{r},\mathbf{r}',\omega'\right)G_{p'p''}\left(\mathbf{r}',\mathbf{r}'',\omega''\right)J_{p'',0}\left(\mathbf{r}'',\omega''\right)
\label{eq:Jp_2}
\end{equation}

\section{Generalized fluctuation-dissipation theorem}

Using the same technique, a generalized form of the fluctuation-dissipation theorem can be written into iterative form as follows

\begin{equation}
\left\langle J_{p}^{*}\left(\mathbf{r}_{1},\omega_{1}\right)J_{q}\left(\mathbf{r}_{2},\omega_{2}\right)\right\rangle =\sum_{m,n=0}^{\infty}\left\langle J_{p,m}^{*}\left(\mathbf{r}_{1},\omega_{1}\right)J_{q,n}\left(\mathbf{r}_{2},\omega_{2}\right)\right\rangle 
\label{eq:FDT_series}
\end{equation}

\noindent where the brackets denote a thermal ensemble average. All current components in (\ref{eq:Jp_series})-(\ref{eq:Jp_2}) are written in terms of the zeroth-order current components. Consequently, all current density correlations in (\ref{eq:FDT_series}) can be written as a function of the zeroth-order current correlations. Since the zeroth-order currents correspond to those for a non-time-modulated body at thermal equilibrium at temperature $T$, one could hypothesize a current correlation of the following form:

\begin{equation}
\left\langle J_{p,0}^{*}\left(\mathbf{r}_{1},\omega_{1}\right)
J_{q,0}\left(\mathbf{r}_{2},\omega_{2}\right)\right\rangle 
=J_T^2\,\,\,\delta_{pq}\,\delta\left(\mathbf{r}_{1}-\mathbf{r}_{2}\right)\,\delta\left(\omega_{1}-\omega_{2}\right)
\label{eq:J0J0}
\end{equation}

Equation\,(\ref{eq:J0J0}) minimally assumes that a body at thermal equilibrium is characterized by fluctuating currents that are uncorrelated in space $\delta\left(\mathbf{r}_{1}-\mathbf{r}_{2}\right)$,
frequency $\delta\left(\omega_{1}-\omega_{2}\right)$, 
and polarization $\delta_{pq}$.
In addition, the strength of such correlations is given by the prefactor $J_T^2$, whose choice will be justified in the following sections. 

In any event, by using (\ref{eq:J0J0}) and (\ref{eq:Jp_series})-(\ref{eq:Jp_2}) it is possible to write the first terms of the generalized current density correlations in iterative form as follows:

\begin{equation}
\left\langle J_{p,1}^{*}\left(\mathbf{r}_{1},\omega_{1}\right)J_{q,0}\left(\mathbf{r}_{2},\omega_{2}\right)\right\rangle =\frac{J_{T}^{2}}{2\pi}\,\frac{\omega_{1}}{c}\frac{\omega_{2}}{c}\,\chi_{TM}^{*}\left(\mathbf{r}_{1},\omega_{1},-\omega_{2}\right)G_{pq}^{*}\left(\mathbf{r}_{1},\mathbf{r}_{2},\omega_{2}\right)
\label{eq:J1J0}
\end{equation}

\[
\left\langle J_{p,1}^{*}\left(\mathbf{r}_{1},\omega_{1}\right)J_{q,1}\left(\mathbf{r}_{2},\omega_{2}\right)\right\rangle =\frac{J_{T}^{2}}{2\pi}\,\sum_{p'}\int\frac{d\omega_{1}'}{2\pi}\int dV_{1}'\,\frac{\omega_{1}}{c}\left(\frac{\omega_{1}'}{c}\right)^{2}\frac{\omega_{2}}{c}
\]

\begin{equation}
\chi_{TM}^{*}\left(\mathbf{r}_{1},\omega_{1},-\omega_{1}'\right)G_{pp'}^{*}\left(\mathbf{r}_{1},\mathbf{r}_{1}',\omega_{1}'\right)\chi_{TM}\left(\mathbf{r}_{2},\omega_{2},-\omega_{1}'\right)G_{qp'}\left(\mathbf{r}_{2},\mathbf{r}_{1}',\omega_{1}'\right)
\label{eq:J1J1}
\end{equation}

\[
\left\langle J_{p,2}^{*}\left(\mathbf{r}_{1},\omega_{1}\right)J_{q,0}\left(\mathbf{r}_{2},\omega_{2}\right)\right\rangle =\frac{J_{T}^{2}}{2\pi}\,\sum_{p',p''}\int\frac{d\omega_{1}'}{2\pi}\,\int dV_{1}'\,\frac{\omega_{1}}{c}\left(\frac{\omega_{1}'}{c}\right)^{2}\frac{\omega_{2}}{c}\,
\]

\begin{equation}
\chi_{TM}^{*}\left(\mathbf{r}_{1},\omega_{1},-\omega_{1}'\right)\chi_{TM}^{*}\left(\mathbf{r}_{1}',\omega_{1}',-\omega_{2}\right)G_{pp'}^{*}\left(\mathbf{r}_{1},\mathbf{r}_{1}',\omega_{1}'\right)G_{p'q}^{*}\left(\mathbf{r}_{1}',\mathbf{r}_{2},\omega_{2}\right)
\label{eq:J2J0}
\end{equation}

An important observation from Eqs.\,(\ref{eq:J1J0})-(\ref{eq:J2J0}) is that the semiclassical theory correctly predicts the existence of nonlocal correlations in space, frequency and position for the fluctuating currents. Therefore, this effect can be directly ascribed to the fact that temporal modulation moves the system outside the equilibrium, and no quantum theory is required for the existence of such correlations. At the same time, the specific values of the correlations could be different, as it will be discussed in Section\,\ref{sec:Discussion}. Finally, we note that the correlations discussed here take place at the level of current densities. Thermal fields from non-time-modulated bodies at thermal correlations can be engineered to present nontrivial spatial and polarization correlations \cite{Joulain2005surface}. However, such correlations are observed at the field level, and arise from the coupling of uncorrelated currents to delocalized optical modes. By contrast, the nonlocal correlations in Eqs.\,(\ref{eq:J1J0})-(\ref{eq:J2J0}) take place at the current level, an effect that has not been predicted for a non-time-modulated body. 

\section{Thermal emission spectrum}

Once the statistical properties of the fluctuating currents are known, the thermal emission spectrum can be calculated as the result of the fields radiated by such currents

\[
S\left(\omega\right)\propto\left\langle \mathbf{E}^{*}\left(\mathbf{r},\omega\right)\cdot\mathbf{E}\left(\mathbf{r},\omega\right)\right\rangle =
\]

\begin{equation}
=\omega^{2}\mu_{0}^{2}\,\sum_{s,p,q}\,\int dV_{1}\,\int dV_{2}\,G_{sp}^{*}\left(\mathbf{r},\mathbf{r}_{1},\omega\right)G_{sq}\left(\mathbf{r},\mathbf{r}_{2},\omega\right)\left\langle J_{p}^{*}\left(\mathbf{r}_{1},\omega\right)J_{q}\left(\mathbf{r}_{2},\omega\right)\right\rangle
\label{eq:S} 
\end{equation}

Following the results of previous sections, this directly implies that the emission spectrum can be written as a series expansion:

\begin{equation}
S\left(\mathbf{r},\omega\right)=\sum_{m,n=0}^{\infty}\,S_{mn}\left(\mathbf{r},\omega\right)
\label{eq:S_series}
\end{equation}

\noindent where

\[
S_{mn}\left(\mathbf{r},\omega\right)=\omega^{2}\mu_{0}^{2}\,\sum_{s,p,q}\,\int dV_{1}\,\int dV_{2}
\]

\begin{equation}
\,G_{sp}^{*}\left(\mathbf{r},\mathbf{r}_{1},\omega\right)G_{sq}\left(\mathbf{r},\mathbf{r}_{2},\omega\right)\left\langle J_{p,m}^{*}\left(\mathbf{r}_{1},\omega\right)J_{q,n}\left(\mathbf{r}_{2},\omega\right)\right\rangle 
\label{eq:Smn}
\end{equation}

Then, by using (\ref{eq:J0J0})-(\ref{eq:J2J0}), the first terms (up to the second order) of the series can be explicitly written as follows

\begin{equation}
S_{00}\left(\mathbf{r},\omega\right)=\omega^{2}\mu_{0}^{2}\,J_{T}^{2}\,\sum_{s,p}\int dV_{1}\,\left|G_{sp}\left(\mathbf{r},\mathbf{r}_{1},\omega\right)\right|^{2}
\label{eq:S_00}
\end{equation}

\[
S_{10}\left(\mathbf{r},\omega\right)=\omega^{2}\mu_{0}^{2}\,\frac{J_{T}^{2}\,}{2\pi}\,\frac{\omega^{2}}{c^{2}}\,\sum_{s,p,q}\,\int dV_{1}\,\int dV_{2}\,\chi_{TM}^{*}\left(\mathbf{r}_{1},\omega,-\omega\right)
\]

\begin{equation}
G_{sp}^{*}\left(\mathbf{r},\mathbf{r}_{1},\omega\right)G_{pq}^{*}\left(\mathbf{r}_{1},\mathbf{r}_{2},\omega\right)G_{sq}\left(\mathbf{r},\mathbf{r}_{2},\omega\right)
\label{eq:S_10}
\end{equation}

\[
S_{11}\left(\mathbf{r},\omega\right)=\omega^{2}\mu_{0}^{2}\,\frac{J_{T}^{2}}{2\pi}\,\sum_{s,p,q,p'}\,\int dV_{1}\,\int dV_{2}\,\int dV_{1}'\,
\]

\[
\int\frac{d\omega_{1}'}{2\pi}\,\frac{\omega^{2}}{c^{2}}\left(\frac{\omega_{1}'}{c}\right)^{2}\chi_{TM}^{*}\left(\mathbf{r}_{1},\omega,-\omega_{1}'\right)\chi_{TM}\left(\mathbf{r}_{2},\omega,-\omega_{1}'\right)
\]

\begin{equation}
G_{sp}^{*}\left(\mathbf{r},\mathbf{r}_{1},\omega\right)G_{pp'}^{*}\left(\mathbf{r}_{1},\mathbf{r}_{1}',\omega_{1}'\right)G_{sq}\left(\mathbf{r},\mathbf{r}_{2},\omega\right)G_{qp'}\left(\mathbf{r}_{2},\mathbf{r}_{1}',\omega_{1}'\right)
\label{eq:S_11}
\end{equation}

\[
S_{20}\left(\mathbf{r},\omega\right)=\omega^{2}\mu_{0}^{2}\,\frac{J_{T}^{2}}{2\pi}\,\sum_{s,p,q,p'}\,\int dV_{1}\int dV_{2}\int dV_{1}'
\]

\[
\int\frac{d\omega_{1}'}{2\pi}\,\frac{\omega^{2}}{c^{2}}\left(\frac{\omega_{1}'}{c}\right)^{2}\chi_{TM}^{*}\left(\mathbf{r}_{1},\omega,-\omega_{1}'\right)\chi_{TM}^{*}\left(\mathbf{r}_{1}',\omega_{1}',-\omega\right)
\]

\begin{equation}
G_{sp}^{*}\left(\mathbf{r},\mathbf{r}_{1},\omega\right)G_{pp'}^{*}\left(\mathbf{r}_{1},\mathbf{r}_{1}',\omega_{1}'\right)G_{p'q}^{*}\left(\mathbf{r}_{1}',\mathbf{r}_{2},\omega\right)G_{sq}\left(\mathbf{r},\mathbf{r}_{2},\omega\right)
\label{eq:S_20}
\end{equation}

The emission spectrum derived in this section can be applied to a wide range of configurations, including arbitrary geometries, as well as arbitrary temporal modulation profiles. Although equations (\ref{eq:S_00})-(\ref{eq:S_20}) might seem complicated to evaluate, it is worth highlighting that all the numerical complexity in the calculation has been transformed into the calculations of dyadic Green's function for a non-time-modulated systems.

\section{Periodic time modulations}
 
As a representative example of temporal modulations, we consider a body whose properties are periodically modulated in time, which is a very popular choice for time-varying media, particularly within the context of photonic time crystals \cite{Lyubarov2022amplified,Dikopoltsev2022light}, as well as  in the investigation of thermal emission effects \cite{Vazquez2023incandescent,Yu2023manipulating,Yu2023time,Buddhiraju2020photonic}. To this end, we consider a time-modulated susceptibility with amplitude $\triangle\chi$ and frequency of modulation $\Omega$:

\begin{equation}
\chi_{TM}\left(\mathbf{r},t,t'\right)=
\triangle\chi\,\delta\left(t-t'\right)\,\mathrm{sin}\,\left(\Omega t'\right)
\label{eq:Chi_th}
\end{equation}

\noindent and, accordingly, with a two-frequency spectrum

\begin{equation}
\chi_{TM}\left(\mathbf{r},\omega,-\omega'\right)=-i\pi\triangle\chi\,\left\{ \delta\left(\omega+\Omega-\omega'\right)-\delta\left(\omega-\Omega-\omega'\right)\right\} 
\label{eq:Chi_th_w}
\end{equation}

By introducing (\ref{eq:Chi_th_w}) into (\ref{eq:J1J0})-(\ref{eq:J2J0}), we find that the generalized current correlations for the particular case of a time-harmonic modulation can be written as
\[
\left\langle J_{p,1}^{*}\left(\mathbf{r}_{1},\omega_{1}\right)
J_{q,0}\left(\mathbf{r}_{2},\omega_{2}\right)\right\rangle =
\] 

\begin{equation}
i\pi\triangle\chi\,\,\frac{J_{T}^{2}}{2\pi}\,\frac{\omega_{1}}{c}\frac{\omega_{2}}{c}\,\left\{ \delta\left(\omega_{1}+\Omega-\omega_{2}\right)-\delta\left(\omega_{1}-\Omega-\omega_{2}\right)\right\} G_{pq}^{*}\left(\mathbf{r}_{1},\mathbf{r}_{2},\omega_{2}\right)
\end{equation}

\[
\left\langle J_{p,1}^{*}\left(\mathbf{r}_{1},\omega_{1}\right)J_{q,1}\left(\mathbf{r}_{2},\omega_{2}\right)\right\rangle =\pi^{2}\triangle\chi^{2}\,\frac{J_{T}^{2}}{2\pi}\,\sum_{p'}\int\frac{d\omega_{1}'}{2\pi}\int dV_{1}'\,\frac{\omega_{1}}{c}\left(\frac{\omega_{1}'}{c}\right)^{2}\frac{\omega_{2}}{c}
\]

\[
\left\{ \delta\left(\omega_{1}+\Omega-\omega_{1}'\right)-\delta\left(\omega_{1}-\Omega-\omega_{1}'\right)\right\} 
\left\{ \delta\left(\omega_{2}+\Omega-\omega_{1}'\right)-\delta\left(\omega_{2}-\Omega-\omega_{1}'\right)\right\} 
\]

\begin{equation}
G_{pp'}^{*}\left(\mathbf{r}_{1},\mathbf{r}_{1}',\omega_{1}'\right)G_{qp'}\left(\mathbf{r}_{2},\mathbf{r}_{1}',\omega_{1}'\right)
\end{equation}

\[
\left\langle J_{p,2}^{*}\left(\mathbf{r}_{1},\omega_{1}\right)J_{q,0}\left(\mathbf{r}_{2},\omega_{2}\right)\right\rangle =-\pi^{2}\frac{J_{T}^{2}}{2\pi}\,\sum_{p'}\int\frac{d\omega_{1}'}{2\pi}\,\int dV_{1}'\,\frac{\omega_{1}}{c}\left(\frac{\omega_{1}'}{c}\right)^{2}\frac{\omega_{2}}{c}\,
\]

\[
\left\{ \delta\left(\omega_{1}+\Omega-\omega_{1}'\right)-\delta\left(\omega_{1}-\Omega-\omega_{1}'\right)\right\} \left\{ \delta\left(\omega'_{1}+\Omega-\omega_{2}\right)-\delta\left(\omega'_{1}-\Omega-\omega_{2}\right)\right\} 
\]

\begin{equation}
G_{pp'}^{*}\left(\mathbf{r}_{1},\mathbf{r}_{1}',\omega_{1}'\right)G_{p'q}^{*}\left(\mathbf{r}_{1}',\mathbf{r}_{2},\omega_{2}\right)
\end{equation}

Similarly, by introducing (\ref{eq:Chi_th_w}) into (\ref{eq:S_10})-(\ref{eq:S_20}), the contribution to the emission spectrum up to the second order can be written as follows

\begin{equation}
S_{00}\left(\mathbf{r},\omega\right)=\omega^{2}\mu_{0}^{2}\,J_{T}^{2}\,\sum_{s,p}\int dV_{1}\,\left|G_{sp}\left(\mathbf{r},\mathbf{r}_{1},\omega\right)\right|^{2}
\label{eq:S_00_th}
\end{equation}

\begin{equation}
S_{10}\left(\mathbf{r},\omega\right)=0
\label{eq:S_10_th}
\end{equation}

\[
S_{11}\left(\mathbf{r},\omega\right)=J_{T}^{2}\,\frac{\triangle\chi^{2}\omega^{2}\mu_{0}^{2}}{4}\,\sum_{s,p,q,p'}\,\int dV_{1}\,\int dV_{2}\,\int dV_{1}'\,\,\frac{\omega^{2}}{c^{2}}G_{sp}^{*}\left(\mathbf{r},\mathbf{r}_{1},\omega\right)G_{sq}\left(\mathbf{r},\mathbf{r}_{2},\omega\right)
\]

\[
\left\{ \left(\frac{\omega+\Omega}{c}\right)^{2}G_{pp'}^{*}\left(\mathbf{r}_{1},\mathbf{r}_{1}',\omega+\Omega\right)G_{qp'}\left(\mathbf{r}_{2},\mathbf{r}_{1}',\omega-\Omega\right)+\right.
\]

\begin{equation}
\left.+\left(\frac{\omega-\Omega}{c}\right)^{2}G_{pp'}^{*}\left(\mathbf{r}_{1},\mathbf{r}_{1}',\omega-\Omega\right)G_{qp'}\left(\mathbf{r}_{2},\mathbf{r}_{1}',\omega-\Omega\right)\right\}
\label{eq:S11_th}
\end{equation}

\[
S_{20}\left(\mathbf{r},\omega\right)=\pi^{2}\triangle\chi^{2}\omega^{2}\mu_{0}^{2}\,\frac{J_{T}^{2}}{2\pi}\,\sum_{s,p,q,p'}\,\int dV_{1}\int dV_{2}\int dV_{1}'\frac{1}{2\pi}\frac{\omega^{2}}{c^{2}}
\]

\[
G_{sp}^{*}\left(\mathbf{r},\mathbf{r}_{1},\omega\right)G_{p'q}^{*}\left(\mathbf{r}_{1}',\mathbf{r}_{2},\omega\right)G_{sq}\left(\mathbf{r},\mathbf{r}_{2},\omega\right)
\]

\begin{equation}
\left\{ \left(\frac{\omega+\Omega}{c}\right)^{2}G_{pp'}^{*}\left(\mathbf{r}_{1},\mathbf{r}_{1}',\omega+\Omega\right)+\left(\frac{\omega-\Omega}{c}\right)^{2}G_{pp'}^{*}\left(\mathbf{r}_{1},\mathbf{r}_{1}',\omega-\Omega\right)\right\} 
\label{eq:S20_th}
\end{equation}

Here, we note that, in accordance to the quantum theory \cite{Vazquez2023incandescent}, the semiclassical theory correctly predicts that the $S_{10}$ first-order contribution to the emission spectrum is zero. The quantum theory establishes that the response of time-varying media is essentially composed by two modes squeezing transformations \cite{Liberal2023quantum,Vazquez2022shaping,Ganfornina2023quantum}, where photons are produced in pairs. Therefore, one might be tempted to conclude that the zero contribution of the $S_{10}$ spectrum relates to photon pair generation. However, the semiclassical theory highlights that this is not the case. The effect directly stems from the uncorrelated nature of the zeroth-order contributions, and it is specific to time-harmonic modulations.

\section{Discussion and conclusions \label{sec:Discussion}}

Up to this point, we have developed a semiclassical theory of thermal emission from time-varying media, based on the usual framework of fluctuational electrodynamics. Within this framework, thermal emission is described as the fields radiated by fluctuating electromagnetic currents with assumed statistical properties. The semiclassical theory correctly recovers important aspects of thermal emission from time-varying media, such as the presence of nonlocal spatial, frequency, and polarization correlations at the current level, as well as the zero contribution from the $S_{01}$ first order spectrum for time-harmonic modulation. 

The only remaining piece of the theory is to specify the value of the strength of the fluctuating currents, $J_T^2$. Following the usual application of fluctuational electrodynamics, the current strength would be the following \cite{Joulain2005surface,Rytov1989principles}:

\begin{equation}
J_T^2
=4\pi\omega\varepsilon_{0}\,\mathrm{Im}\left[\chi_{S}\left(\mathbf{r},\omega\right)\right]\Theta\left(\omega,T\right)
\end{equation}

By comparing the resulting theory with the quantum formalism developed in \cite{Vazquez2023incandescent}, we find that both theories are almost identical, except that the semiclassical theory fails to predict the occurrence of quantum vacuum amplifications effects. Therefore, the predictions from the semiclassical theory should be accurate for sufficiently high temperatures, but it will fail to predict the presence of emission in the zero temperature limit, and it will be inaccurate for a intermediate range of temperatures. 

\begin{figure}[t!]
\centering
\includegraphics[width=0.9\textwidth]{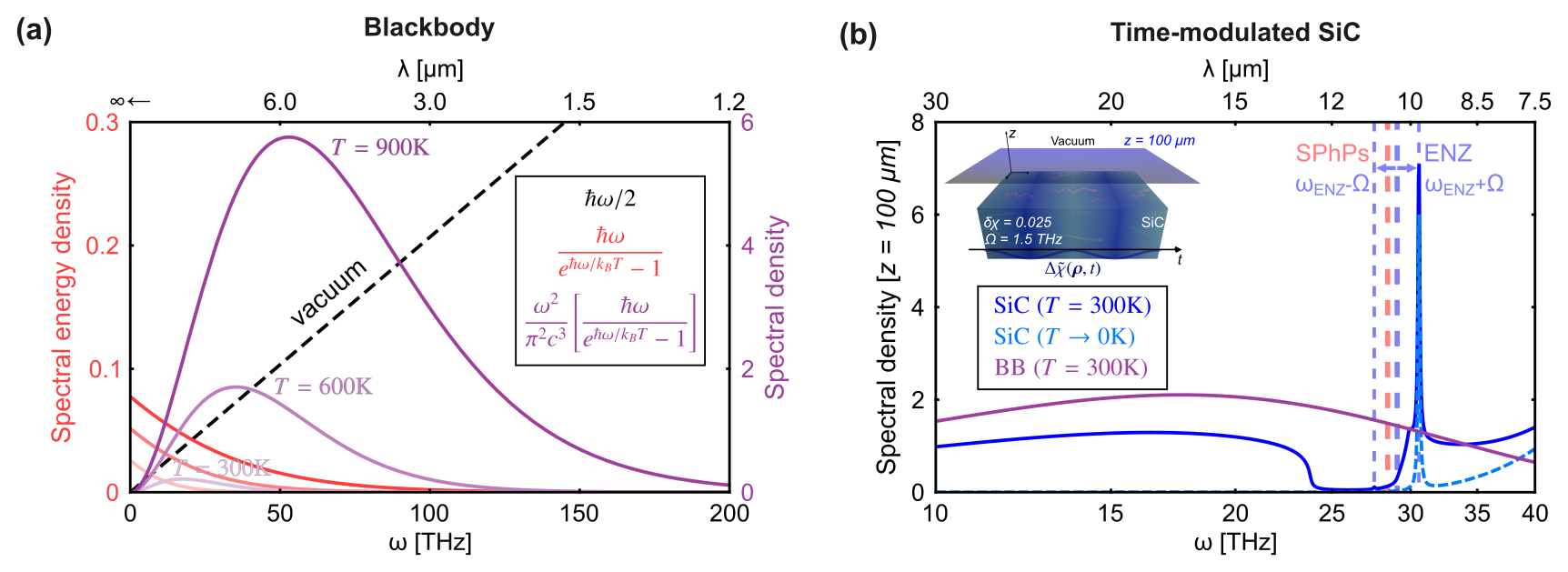}
\caption{ { \bf Vacuum contribution to thermal emission from time-varying media.} (a) Comparison between the vacuum energy, $\hslash\omega/2$, thermal energy, $\Theta\left(\omega,T\right)=\hslash\omega/\left[{\rm exp}\left(\hslash\omega/k_{B}T\right)-1\right]$, and blackbody radiation, $\frac{\omega^2}{\pi^2 c^3}\,\,\Theta\left(\omega,T\right)$, as a function of frequency/wavelength. (b) Comparison between the room temperature (300\,K) and zero-temperature limit (0\,K) thermal emission spectra from a silicon carbide (SiC) substrate temporally modulated with a time-harmonic profile: 
$\chi_{TM}(t,\tau) = \delta\chi\,\,\delta(t-\tau)\sin(\Omega t)$.}
\label{fig:Vacuum_contribution}
\end{figure}

A reasonable question to ask is how critical capturing quantum vacuum amplification effects can be. After all, similar phenomena, such as the dynamical Casimir \cite{Dodonov2020fifty,Wilson2011observation} and Unruh \cite{Crispino2008unruh} effects, are considered exotic and extremely hard to observe. Despite this, we argue that the ultrafast modulation in time-varying media will lead to quantum vacuum amplification effects comparable or larger than thermal emission, even at ambient temperature. 
This effect can be intuitively understood by comparing the spectral energy density of thermal and quantum vacuum fluctuations. As shown in Fig.\,\ref{fig:Vacuum_contribution}(a), the vacuum energy spectrum $\hslash\omega/2$ meets the thermal energy spectrum 
$\Theta\left(\omega,T\right)=\hslash\omega/\left[{\rm exp}\left(\hslash\omega/k_{B}T\right)-1\right]$
at MIR frequencies. Moreover, such meeting points take place even at wavelengths longer than the peak of thermal radiation from a blackbody, 
$\frac{\omega^2}{\pi^2 c^3}\,\,\Theta\left(\omega,T\right)$.
Therefore, when met with an amplifying mechanism such as time-modulation, they can be expected to produce equally important contributions, even at wavelengths comparable and/or longer than those in which thermal emission is maximized. 

Such intuition is confirmed in the calculations reported in Fig.\,\ref{fig:Vacuum_contribution}(b), which depicts a comparison of the spectral energy density radiated by a silicon carbide (SiC) substrate at room temperature (300\,K), and in the zero-temperature limit (0\,K). For this calculation silicon carbide was modelled with Lorentzian dielectric permittivity:
$\varepsilon_{SiC}=
\varepsilon_{\infty}
(\omega^2 - \omega_{ENZ}^2 + i\omega\omega_c)/
(\omega^2 - \omega_{0}^2 + i\omega\omega_c)$, with 
$\varepsilon_{\infty}=6.7$,
$\omega_{ENZ}=2\pi\cdot 29.1\cdot 10^{12}\,$rad/s,
$\omega_{ENZ}=2\pi\cdot 23.8\cdot 10^{12}\,$rad/s,
$\omega_c=2\pi\cdot 0.14\cdot 10^{12}\,$rad/s.
In addition, we assumed a time-modulated susceptibility with a time-harmonic profile: 
$\chi_{TM}(t,\tau) = \delta\chi\,\,\delta(t-\tau)\sin(\Omega t)$, with $ \delta\chi=0.025$ and $\Omega = 2\pi\cdot 1.5\cdot 10^{12}\,$rad/s. The calculations confirm that the emission in the zero-temperature limit, given exclusively by quantum vacuum amplification, is comparable to thermal emission at room temperature. In fact, the time-modulated part of the emission is dominated by the quantum vacuum amplification contribution. 

From this perspective, semiclassical theories would fail to accurately predict thermal emission from time-varying media, since capturing quantum vacuum amplification is critical for time-modulated systems at MIR frequencies. As a solution, one might attempt to fix the semiclassical theory so that it phenomenologically includes vacuum fluctuations. Indeed, previous works on thermal emission based on fluctuational electrodynamics have considered the use of symmetrized current density correlations  \cite{Joulain2005surface}, which would include vacuum fluctuations. In this manner, the strength of the fluctuating currents would be given by 

\[
\left\langle 
J_{p,0}^{*}\left(\mathbf{r}_{1},\omega_{1}\right)
J_{q,0}\left(\mathbf{r}_{2},\omega_{2}\right)
\right\rangle
\]

\begin{equation}
 =
\frac{1}{2}\,\left\langle 
\widehat{J}_{p,0}^{\dagger}\left(\mathbf{r}_{1},\omega_{1}\right)
\widehat{J}_{q,0}\left(\mathbf{r}_{2},\omega_{2}\right)
+
\widehat{J}_{p,0}\left(\mathbf{r}_{1},\omega_{1}\right)
\widehat{J}_{q,0}^{\dagger}\left(\mathbf{r}_{2},\omega_{2}\right)
\right\rangle 
\end{equation}

\begin{equation}
J_T^2=4\pi\omega\varepsilon_{0}\,\mathrm{Im}\left[\chi_{S}\left(\mathbf{r},\omega\right)\right]\left(\Theta\left(\omega,T\right)+\frac{\hslash\omega}{2}\right)
\end{equation}

This approach can be intuitively understood as modeling vacuum fluctuations as a background classical noise. It successfully predicts quantum vacuum amplification effects and emission at zero temperature, but it generates additional problems. First, we note that it predicts a vacuum contribution to the emission even from a non-time-modulated body. This problem can be solved by noting that such contribution would be canceled out by the vacuum contribution from the environment. In other words, both the body and the background are in a ``vacuum equilibrium", with equal contributions leading to a zero net power flow. Therefore, it can be justified to drop the $\frac{\hslash\omega}{2}$ contribution for the $S_{00}$ spectrum, since it would not lead to a measurable power transfer. 

However, an additional problem remains. In particular, we note that the semiclassical theory predicts the same factor $J_T^2$ for all higher order terms, while the quantum theory does not. For example, the quantum theory predicts that there is no vacuum contribution for the $S_{20}$ spectrum, while there is one for the $S_{11}$ spectrum. However, the presented semiclassical theory based on fluctuational electrodynamics cannot make such distinction, since all $S_{mn}$ spectra are multiplied by the same prefactor $J_T^2$. The reason for the remaining disagreement is that such differences emerge from the nontrivial commutation rules of the current operators, purely pertaining to the quantum formalism, which cannot be easily incorporated into the semiclassical theory. This result highlights the difficulty of modeling quantum vacuum fluctuations as an additional classical noise. 

At the same time, we note that one could choose a different $J_{mn}^2$ fluctuating current strength factor for each $S_{mn}$ spectrum, thus fitting the semiclassical theory to the results of the quantum formalism. For the first three terms, the choices should be the following:

\begin{equation}
J_{T,00}^2=4\pi\omega\varepsilon_{0}\,\mathrm{Im}\left[\chi_{S}\left(\mathbf{r},\omega\right)\right]\Theta\left(\omega,T\right)
\end{equation}

\begin{equation}
J_{T,20}^2=4\pi\omega\varepsilon_{0}\,\mathrm{Im}\left[\chi_{S}\left(\mathbf{r},\omega\right)\right]\Theta\left(\omega,T\right)
\end{equation}

\begin{equation}
J_{T,11}^2=4\pi\omega\varepsilon_{0}\,\mathrm{Im}\left[\chi_{S}\left(\mathbf{r},\omega\right)\right]\left(\Theta\left(\omega,T\right)+\frac{\hslash\omega}{2}\right)
\end{equation}

With the use of these current strength coefficients accompanying each spectrum, it is possible to calculate thermal emission from time-varying media with the use of classical tools. The prize to pay is that all physical intuition behind such current density correlations is arguably lost. Moreover, an additional shortcoming of semiclassical formalism is that they are unable to predict the photon statistics of the generated light, and the nontrivial correlations of light emission from time-varying media, which is a subject of current interest \cite{Liberal2023quantum,Vazquez2022shaping,Ganfornina2023quantum}.

As a final note, we remark that we have derived a semiclassical theory based on the standard form of fluctuational electrodynamics, which is nowadays the most common tool for the design of thermal emitters. However, it cannot be ruled out that there could be other semiclassical theories based on heuristic arguments that could provide a fit to the quantum theory. 

\section*{References}
\bibliographystyle{iopart-num}
\bibliography{library}

\end{document}